\magnification=\magstep1
\pretolerance 2000
\baselineskip=12pt
\catcode`@=11
\def\vfootnote#1{\insert\footins\bgroup\baselineskip=20pt
  \interlinepenalty\interfootnotelinepenalty
  \splittopskip\ht\strutbox 
  \splitmaxdepth\dp\strutbox \floatingpenalty\@MM
  \leftskip\z@skip \rightskip\z@skip \spaceskip\z@skip \xspaceskip\z@skip
  \textindent{#1}\footstrut\futurelet\next\fo@t}
\skip\footins 20pt plus4pt minus4pt
\def\footstrut{\vbox to2\splittopskip{}}
\catcode`@=12
\def\folio{\ifnum\pageno=0\else\ifnum\pageno<0 \romannumeral-\pageno
\else\number\pageno \fi\fi}
\def\buildurel#1\under#2{\mathrel{\mathop{\kern0pt #2}\limits_{#1}}}
\vglue 24pt plus 12pt minus 12pt
\bigskip
\centerline {\bf  Current-Voltage Characteristic of a Partially}
\centerline {\bf  Ionized Plasma in Cylindrical Geometry}
\bigskip
\bigskip
\centerline {\bf by}
\centerline {Joel L. Lebowitz and Alexander Rokhlenko}
\centerline {Department of Mathematics}
\centerline {Hill Center, Busch Campus}
\centerline {Rutgers University}
\centerline {New Brunswick, NJ 08903}
\bigskip
\bigskip\bf
\centerline {ABSTRACT}
\medskip\rm
    The properties of a partially ionized plasma in a long cylindrical
tube subject to a uniform axial electric field are investigated. The
plasma is maintained by an external ionizing source balanced by bulk
and surface recombinations. Collisions between neutrals, whose density
greatly exceeds the density of charged particles, and of neutrals with
ions are sufficiently effective for their velocity distribution to be
close to a Maxwellian with the same uniform temperature, independent
of the external field. The behavior of the plasma is described by a
collisional two-fluid scheme with charge neutrality in the interior of
the tube.  Approximate nonlinear equations for the hydrodynamical
moments are obtained from a Boltzmann equation in which
electron-neutral, electron-ion and electron-electron collisions are
all important. It is found that under certain circumstances the
current, and the temperature of the electrons undergo a drastic
change, with hysteresis, as the electric field is varied.
\bigskip\bf
I. Introduction
\medskip\rm
Instabilities are ubiquitous in strongly ionized plasmas. They
dominate the behavior of such systems and their study forms the core
of the subject. The origin of the instabilities lies in the nature of
the plasma interactions: on the one hand they are long range and thus
can produce strong cooperative effects and on the other hand they
become 'weaker' locally at high energies (or temperatures) as
manifested by the decrease of the Coulomb cross section with energy
rise$^1$. The situation is different in weakly ionized cold plasmas,
systems which have attracted much attention recently$^{2,3}$. In such
systems collective phenomena play a smaller role and instabilities are
less common.  Nevertheless there are cooperative phenomena in these
systems too, which, as we have shown earlier$^{4,5}$ for homogeneous
idealized systems and will show here for more realistic laboratory
situations, can lead to dramatic abrupt changes in the state of the
plasma, when such systems are driven by external fields.

There is a large literature on the behavior of plasma in a cylindrical
geometry (see for example Refs.6,7,8,9). The reason for considering
this system again here is that we are interested in a regime in which
the electrons are colder than in the self-sustained discharge plasmas
generally investigated. The plasma has a sufficiently high degree of
ionization to make the electron- electron and electron-ion collisions
important. The condition of low electron temperature permits us to
neglect inelastic collisions but requires that we include volume
recombination effects.

The phenomenon in which we are most interested is a rapid, essentially
abrupt, change in the electron temperature and current as the external
electric field crosses a certain critical value which depends on the
neutral--ion temperature, degree of ionization, etc. The phenomenon is
related to the well known runaway effect in fully ionized plasmas
caused by the decrease of the electron-ion collision cross section as
the external field increases the electron energy$^{10,11,12}$. In the
partially ionized gas the presence of the neutrals prevents such a
runaway. In fact, if one neglects collisions between the electrons
(e-e), as is done in the swarm approximation$^{13}$, valid when the
degree of ionization is sufficiently small, then the stationary
distribution (in the absence of inelastic collisions) will be of the
Druyvesteyn form$^{14}$ for which the current and electron mean energy
are smooth functions of the external field. The situation is similar
when the size of the plasma is smaller than the energy relaxation
length of the electrons$^{9}$. There are, however, other regimes, even
in weakly ionized plasmas, where the effect of the e-e collisions is
sufficient to keep the electron distribution close to a
Maxwellian$^{10,4}$. Such conditions can lead to cooperative abrupt
changes in the temperature and drift velocity of electrons as the
external field is varied. The origin of the phenomenon is made clearer
if one starts with a kinetic approach rather than with a macroscopic
description.

While it is possible, even likely, that effects related to those
discussed here have already been observed indirectly in the behavior
of discharges it would be useful for both theoretical and practical
reasons to have experiments in which the parameters can be controlled,
so as to study the phenomena in a quantitative way. We expect that the
transition will be seen as hysteresis in the current--voltage (I-V)
characteristic of the plasma when the external electric field is
slowly varying in time$^{4,5}$.

We shall consider an experimental arrangement consisting of a weakly
ionized gas in a tube of radius $R$ subjected to a constant external
axial electric field ${\bf E}.$ The plasma is assumed for simplicity
to be produced through uniform ionization inside the tube, by some
external source, at a constant rate $a$. It is balanced by two kinds
of recombination processes: a bulk one and a surface process on the
tube wall. The main bulk recombination for the regime we are
interested in, are$^{10}$ three body processes involving two electrons
plus an ion,
and dissociative recombinations in which a metastable atom-ion complex
recombines with an electron.  The rates depend on the temperatures and
densities of the electrons, ions and neutrals in a rather complicated
way; see Ref.15.
{}For the sake of simplicity we lump the two processes together and 
assume an effective rate of bulk recombination proportional to 
$T^{-3/2}$, where $T$ is the electron temperature. We ignore the 
dependence of this rate on the neutral and ion temperature and on 
the pressure which we keep more or less constant.
The recombination at the wall is also treated phenomenologically. 
In particular we assume that the energy is absorbed by the wall which 
is kept at a fixed temperature, see section 4. 
\medskip
We have in mind here a situation in which the great majority of
neutral atoms are some kind of noble gas to which may, or may not, be
added a small amount of a more easily ionized second species, though
we realize that in the latter case the analysis would be more
complicated$^{16}$.  This will be reflected mainly in the rates of
ionization $a$ and recombination $\gamma$ since we shall always
consider a regime in which the density of electrons $n(r,t)$ is much
lower than that of the neutrals, $N$, but big enough, due to the great
disparity between the electron mass, $m$, and the ion--neutral mass,
$M$, for binary electron-electron collisions to dominate the energy
exchanges in electron-ion and electron-neutral (e-i and e-n)
collisions. This requires$^4$ that $$ {\sigma m\over 2\pi Me^4}(kT)^2
<<{n\over N}<<1.\eqno(1a)$$ Here $k$ is the Boltzmann constant, and
$\sigma$ is the total electron-neutral particle collision cross
section, which is taken to be a constant in our work. Putting in
appropriate values for the parameters on the left side of (1a) gives
(see Ref.4),$$ 1>>n/N >> 8\cdot 10^{-7},\ 4\cdot 10^{-8},\ 1.5\cdot
10^{-8}\eqno(1b)$$ for He, Ne, Ar plasmas respectively, when $kT$ is
approximately 1 eV and it decreases as $T^2$ for colder electrons.
The upper bound relates to the fact that we ignore any collective self
induced electrostatic or magnetic interactions.
\medskip
The ions in our model are assumed to have a uniform temperature,
$T_i$, the same as the neutrals, while their density is $n(r,t)$,
i.e. the plasma is treated as locally quasi-neutral. We assume axial
symmetry and longitudinal homogeneity so the spatial dependence is
only in the radial variable $r\leq R$. The different mobilities of
ions and electrons are compensated by an internally generated radial
ambipolar electric field ${\bf F}(r,t)$. We are thinking of a
quiescent, long positive column which fills the tube$^6.$
\medskip
The reason for considering external rather than field induced
ionization is that in our previous works$^{4,5}$, in which we
considered a spatially homogeneous case with an a priori fixed plasma
density, we found that as we varied the external field $E$, there was
a transition in the electron distribution between regimes of weak and
strong coupling to the ions. (Ref.5 presents a rigorous proof of such
a transition for a greatly simplified model system). This mechanism of
a kinetic transition, which we investigate here in a more realistic
physical situation, requires a low electron temperature and relatively
weak electric field, which seems hard to achieve when the ionization
is produced by the field, see Sec.4.
\medskip
The basic idea in Ref.4 and here is to consider situations in which
the collisions between electrons are strong enough to force their
velocity distribution $f(r,{\bf v}, t)$ to stay close to a Maxwellian
$M_f$ with temperature $T$ and drift velocity ${\bf w}$. The values of
$T$ and ${\bf w}$, which are simply related to the first two velocity
moments of $f$, are then determined by self consistent
``hydrodynamic'' equations, i.e. we evaluate the integrals entering
the time evolution of $T$ and ${\bf w}$ with the help of replacing $f$
by this Maxwellian. {}For the spatially homogeneous case this yields
ordinary differential equations in time for $T$ and ${\bf w}$ which
can be reduced to a couple of transcendental equations for stationary
values $T(E)$ and ${\bf w}(E)$ yielding, in some cases, S-shaped
curves as functions of $E={\bf |E|}$. In the region where $w$
decreases with $E$ the system is unstable. The current and the
electron temperature can therefore be expected to jump upwards
(downwards) when the field increases (decreases).  This will occur at
different values of the field intensity, see Fig.3: in the lower
(upper) part of the loop the electrons will be supercooled
(superheated) creating in this way a hysteresis loop.  The origin of
this behavior lies in a changeover from e-i to e-n coupling as the
dominant factor when the electron energy is increased by the field.

In the situation analyzed in this paper we obtain nonlinear partial
differential equations for $n(r,t),\ T(r,t)$, ${\bf w}(r,t)$ and
$F(r,t).$ Their stationary solutions are the main concern of this
work. This is presented in section 3 following the mathematical
formulation of the problem in section 2.  A discussion of physical
situations where the transition predicted by the two-fluid model used
in this paper might be observed is presented in section 4. This is a
much more restrictive domain than that given by (1), it requires
essentially $n/N \sim 10^{-4}$ while $N \sim 10^{15} cm^{-3}$ for
light noble gases when $T_i$ is close to room temperature. At the
transition the electron temperature jumps within a range of several
$T_i$, staying significantly lower than the usual temperatures in
self-sustaining gas discharges. This determines the experimental set
up we study here. The nature of the various approximations made is
discussed in section 4 and in the Appendix.
\bigskip\bf
II. Mathematical Description
\medskip\rm
   The behavior of the system is described by kinetic equations for
the electron and ion distribution functions $f(r,{\bf v}, t)$ and
$f_i(r,{\bf v}, t)$ normalized to the same density $n(r,t)$. Under the
above assumptions the Boltzmann equation for electrons has the
form$^4$ $$ {\partial f(r,{\bf v}, t)\over \partial t} - {e\over
m}\left [{\bf E} +{{\bf r}\over r}F(r,t)\right ]\cdot {\bf \nabla_v}
f(r,{\bf v}, t)+{\bf v\cdot \nabla_r} f(r,{\bf v}, t)=$$
$$a\psi ({\bf v})- \gamma n f(r,{\bf v,}t)+
{m\over Mv^2}{\partial \over \partial v}
\left \{ (bn+v^4 /l)\left [{\bar f}(r,v,t)+{kT_i\over mv}{\partial
{\bar f}(r,v,t)\over \partial v}\right ]\right \} +$$ 
$${bn\over v^3}\hat L f(r,{\bf v},t)+
{v\over l}[{\bar f}(r,v,t)- f(r,{\bf v}, t)] +Q[f].\eqno(2)$$
In (2) the first two terms on the right side represent the 
ionization and recombination respectively:
$\psi ({\bf v})$ is the normalized distribution function of 
newly born electrons with mean kinetic energy $mv^2_0 /2$,$$
\int \psi ({\bf v})d^3 v=1,\ \ \int {\bf v}\psi ({\bf v})d^3 v 
=0,\ \ \int v^2\psi ({\bf v})d^3 v = v_0^2.$$ The third term with the
prefactor $m/M$, represents the diffusion in the speed of the
electrons due to collisions with ions and neutrals. The effectiveness
of the e-i collisions is proportional to $n$ and is strongly peaked at
small speeds. The constant $b$ is given$^{4,11}$ by $b=4\pi e^4 L/m^2$
($L\sim 10$ is the Coulomb logarithm), while $l=1/N\sigma$ is the mean
free path of electrons in e-n collisions. The next two terms in the
right hand side of (3) represent respectively the effects of e-i and
e-n collisions on the angular parts of electron velocities with$$
\hat L = \Sigma_{\mu,\nu}\ {\partial \over \partial c_{\mu}}
(c^2 \delta _{\mu,\nu} - c_{\mu} c_{\nu} ){\partial \over 
\partial c_{\nu}},\qquad (\mu ,\nu =1,2,3) $$ where ${\bf c = v-W}$
and ${\bf W}$ is the ion drift velocity.  We neglect the difference
between $|{\bf c}|$ and $|{\bf v}|$, in particular in the energy
exchange term in (2), since $|{\bf W}|<<v$. {}For the e-n collisions
we have defined$$ {\bar f}(r,v,t)={1\over 4\pi }\int f(r,{\bf v,}t)\
d\Omega_v$$ as the sphericalized average of $ f(r,{\bf v},t)$.
\medskip
{}For simplicity we have written the e-i and e-n collision terms when
ions and neutrals have approximately the same mass $M$.  In the case
of using an easily ionized impurity these collision terms have to be
modified in an obvious way. However, as long as the density of these
impurities is of order of $n$, and $n/N<< 1$, we do not expect them to
substantially affect our result.  {}Finally the last term $Q[f]$
represents e-e collisions. We do not specify here the exact form of
this term (although we have in mind a Landau form$^{10,11,12}$), since
it conserves the hydrodynamical moments corresponding to density, mean
velocity and kinetic energy which we are going to derive from (3).

We could but do not write down a similar kinetic equation for the
ions.  As already mentioned we assume them to have a distribution
close to a Maxwellian with a uniform temperature $T_i$, equal to that
of the neutrals. The assumption of charge neutrality further implies
that both electrons and ions have the same density$$ n(r,t)=\int
f(r,{\bf v,}t) d^3v =\int f_i(r,{\bf v_i,}t) d^3 v_i$$ and the same
radial drift velocity, as we shall see in the next section. The
maintenance of this quasi-neutrality is the task of the radial
ambipolar field $F(r)$.  The effect of recombination at the wall and
other wall effects will be discussed in the next section.
\bigskip
{\bf III. Hydrodynamic Description}
\medskip
Many properties of the plasma can be expressed in terms of the first
few velocity moments of $f$ and $f_i$. To obtain these we multiply (2)
by $1, {\bf v}, v^2$ and integrate over ${\bf v.}$ This yields
expressions for the time derivatives of the density $n(r,t),$ drift
velocity $\langle {\bf v}\rangle = {\bf w}(r,t)$ and $\langle
v^2\rangle =3kT(r,t)/m +w^2 (r,t)$ which however involve unknown
integrals over $f$. To obtain a closed set of hydrodynamical equations
we follow the procedure in Refs.4,5, which is often used for
collisional plasmas, and compute all integrals over ${\bf v}$ with the
help of the replacement$$ f(r,{\bf v},t)\to M_f(r,{\bf v},t)\eqno(4)$$
where $$M_f(r,{\bf v}, t)=n(r,t)(m/2\pi kT)^{3/2} exp[-m({\bf
v-w})^2/2kT]\eqno(5)$$ is a local Maxwellian with the same $r$ and $t$
dependent first moments, $n, {\bf w},\langle v^2\rangle ,$ as
$f(r,{\bf v},t)$.  This assumption is motivated by the fact that the
local Maxwellian $M_f$ is the unique distribution which makes the e-e
collision term $Q[f]=0$ and should thus be a good approximation (after
some transient time) for situations in which $Q[f]$ is the dominant
term in determining the shape of the distribution function. This
requires, at the minimum, that $n/N$ satisfy (1b) and that $E,\ a$ and
$\gamma$ be small so that the energy exchanges in the e-e collisions
are most important in determining the shape of the energy distribution
of the electrons; the effectiveness of the e-n and e-i collisions
being reduced by the factor $m/M$ which is the physically small
parameter in the problem. Eq.\ (5) can be rigorously justified, as was
shown in Ref.5, for sufficiently idealized models. We believe that the
evaluation of these integrals via $M_f$ remains valid, at least
approximately, in certain realistic situations, see section 4 and
Appendix.
\medskip
Using (4) we obtain four equations for the electron density $n$, the
longitudinal and radial components of {\bf w}, $w_z,w_r$, the
temperature $T$, and the radially directed electric field ${\bf F}.$
In writing down these equations we shall neglect higher order terms in
the drift velocity $w(r)$ as compared to the electron thermal velocity
$u(r)= \sqrt{3kT/m}.$ (This is a good approximation when $m<<M$, see
section 4 and Appendix).
\bigskip\noindent
a) Electron density equation$$ {\partial n\over \partial t}=
-{\partial\over r\partial r}(rnw_r) + a - c\ {n^2\over u^3}.\eqno(6)$$
b) Electron flux in the z-direction$$ {1\over n}{\partial (nw_z)\over
\partial t} + {eE\over m}=-2\sqrt{6
\over\pi}w_z{bn+4u^4/9l\over u^3}-{cnw_z\over u^3}.\eqno(7)$$
c) Electron flux in the radial direction$$
{1\over n}{\partial (nw_r)\over \partial t} + {eF(r,t)\over m}=-
{8\over 9}\sqrt{6\over \pi}w_r{u\over l}-{cnw_r\over u^3}-
{1\over 3n}{\partial \over \partial r}(nu^2).\eqno(8)$$
d) The energy balance equation$$
{1\over 2}{\partial (nu^2)\over \partial t}+n{e\over m}[F(r,t)w_r+
Ew_z]+{5\over 6r}{\partial \over \partial r}(rnw_r u^2 )=$$
$${a v_0^2\over 2} - {cn^2\over 2u} - \sqrt{6\over \pi }{m\over
M}\left (1-{v^2_i\over u^2}\right )n{bn+8u^4/9l\over u}.\eqno(9)$$
Here $$ u^2 (r,t)=3kT(r,t)/m = \langle v^2 \rangle - w^2 (r), \quad
v^2_i =3kT_i/m,$$ and we have set $\gamma$ proportional to $T^{-3/2},\
\gamma =c/u^3$, see Ref.6. The temperature independent parameter $c$
describes qualitatively the rate of the volume recombination. In the
case of dissociative recombination $c$ is typically$^{10,15}$ in the
range $4\times (10^{13} - 10^{15} )cm^6 /sec^4$ and close to the lower
value for pure Helium. The temperature dependence of $\gamma$
represents a combination of both dissociative process and three-body
recombinations. The latter will be most important when the electron
density $n$ is large and $T$ is low. It gives a reasonable rate in the
vicinity of $300^0$ K for the neutral and ion temperatures.
\medskip
A similar procedure with the ion kinetic equation yields (6) for the
zeroth moment because charge neutrality implies the same $n(r,t)$ and
therefore the same $w_r(r,t)$. The ion radial flux is described by the
following equation$$ {1\over n}{\partial (nw_r)\over \partial t} +
{eF(r,t)\over M}=- {8\over 9}\sqrt{6\over \pi}w_r{U\over
l_i}-{cnw_r\over u^3}- {1\over 3n}{\partial \over \partial
r}(nU^2)-{1\over nr} {\partial \over \partial r}(rnw_r^2).\eqno(10)$$
Here $U$ is the ion thermal velocity ($MU^2=3kT_i$) and $l_i$ is the
mean free path in i-n collisions with respect to momentum relaxation.
Unlike in (8) we cannot, in (10), neglect $w_r^2$ compared with $U^2$.
We do however neglect the ionic contribution to the total longitudinal
current and assume that the ions have the same temperature as the
neutrals since they are in a much better thermal contact with the
neutral particles than the electrons. We believe that in our range of
electron temperatures we make only a small error in taking $T_i=const$
and neglecting the ion heating. Dependence of $T_i$ on $r$ can be
easily incorporated by adding one more differential equation similar
to (9).

The nonlinear partial differential Eqs.(6)--(10) have to be solved for
the five unknown functions\ $ F(r), n(r), u(r), w_z (r), w_r (r).$ In
the stationary state the set (6 - 10) can be reduced by the
elimination of $w_z (r)$ and $F(r)$ with the help of (7),(8) and
(10). We obtain
$$w_z (r)= - {1\over 2}\sqrt
{\pi\over 6}{eE\over m}{u^3(r)\over gn(r)+4u^4(r)/9l}\eqno(11)$$ 
and three coupled equations $$
{d\over rdr}(rnw_r) =a - c\ {n^2\over u^3},\eqno(12)$$
$${m\over M}{d\over dr}\left (n{u^2+v_i^2\over 3}\right )+{d\over
 rdr}(rnw_r^2)=
-\left ({8\over 9}\sqrt{6m\over \pi M}{v_i\over l_i}+{cn\over u^3}
\right )nw_r,\eqno(13)$$
$$u(r)n(r){du(r)\over dr}-{1\over 3}u^2(r){dn(r)\over dr}=
{8\over 9l}\sqrt{6\over \pi }nw_r+A(r)/w_r \eqno(14)$$
for $n(r), u(r), w_r(r).$ Here $g=b+{1\over 2}\sqrt{\pi\over 6}c$ and $$
A(r)={1\over 2}\sqrt{\pi \over 6}\left ({eE\over m}\right )^2{u^3(r)
n(r)\over gn(r)+4u^4(r)/9l}+{a\over 2}[v_0^2 -{5\over
3}u^2(r)]+{1\over 3}c{n^2(r)\over u(r)} -$$
$$\sqrt{6\over \pi }{m\over M}
\left [1-{v^2_i \over u^2(r)}\right ]n(r){bn(r)+8u^4(r)/9l\over 
u(r)}.\eqno(15)$$
\medskip
It is well known$^{6,7,8}$ that in the hydrodynamic approximation the
assumption of the charge neutrality leads to singularities near the
plasma boundaries due to the space charge in the plasma sheath caused
by the different thermal velocities of electron and ions and e-i
recombinations near the wall. The singularity manifests itself in
Eqs.(12-14) through the determinant of the linear system for the
derivatives ${dn\over dr},{du\over dr},{dw_r\over dr}$ becoming zero
when the radial drift velocity, at some $r=r'$, reaches the value$$
w_r=\sqrt{k{5T_e+3T_i\over 3M}}\eqno(16)$$ that is close to the
adiabatic sound speed$^{7}$. According to (8) the infinite derivatives
imply an infinite electric field $F(r)$ and hence a breakdown of the
model. Following Refs.7,8 we interpret this to mean that charge
neutrality cannot hold for $r>r'$ and $r'$ marks the boundary of the
plasma sheath.  On the other hand the thickness of the sheath is$^6$
of the order of a few Debye lengths $\lambda _D = \sqrt{kT_i/4\pi ne^2
}\approx 10^{-4}cm$ when $T_i=300 K, \ n\approx 10^{12} cm^{-3}$.
Assuming that $R>>\lambda _D$ we can neglect the difference between
$r'$ and $R$ in Eq.(16) and instead use (16) as the boundary condition
for (12-14), as is proposed by Persson$^7$ and other authors (see
Refs.6,8).

Our task now is to solve (12)--(14) subject to the conditions $$
{dn\over dr}(0)=0,\ \ {du\over dr}(0)=0,\ \ w_r(0)=0\eqno(17)$$ at
$r=0$ and to $w_r(R)$ given by (16) at $r=R$.  The functions $n(r),\
u(r)$ are finite on the tube axis, $r=0$, Eqs.(12), (14), (17)
therefore imply $$ {dw_r\over dr}(0)={1\over
2n(0)}[a-cn^2(0)/u^3(0)]\eqno(18)$$ and $A(0)=0$.  The last
relationship is very important and we rewrite it as$$
\sqrt{\pi \over 6}\left ({eE\over m}\right )^2 =
{gn(0)+4u^4(0)/9l\over u^3 (0)\ n(0)}\Biggl\{                         
2\sqrt{6\over \pi}{m\over M}
\left [1-{v^2_i \over u^2(0)}\right ]n(0){bn(0)+8u^4(0)/9l\over 
u(0)}$$ 
$$- a\left [ v_0^2 -{5\over 3}u^2(0)\right ] -
{2\over 3}c{n^2(0)\over u(0)}\Biggr\}.\eqno(19)$$
\medskip
Solving the set (12-14) with conditions (16-19) we determine the
density and temperature profiles as well as the average temperature
$\bar T(E)$ and the total longitudinal current $I(E)$: $$
\bar T(E)= {2m\over 3kR^2} \int^R_0 u^2 (r)rdr,\qquad 
I(E)= 2\pi e\int^R_0  rn(r)w_z (r)\ dr.\eqno(20)$$
\medskip
When we studied in Ref.4 the spatially homogeneous problem without
ionization and recombination, we had $c=0,\ a =0,\ n=const,\ u=
const$, and (19) served as the equation for the determination of $u$,
or the electron temperature, as a function of $E$. We have a similar
situation here when $l/R <<1$. In this case one can disregard the
recombination on the tube walls and we have again a spatially
homogeneous problem $n=n(0),\ u=u(0)$. The only difference is that we
have now, in virtue of (12),$$ n=\sqrt {a u^3 \over c}.$$ Defining $$
\phi ={9\over 8}\sqrt{\pi\over 6}{lv_i^{-3/2}\over 
\sqrt{ac}}\left ({eE\over m}\right )^2 ,\ \omega = {9 \over 4} 
blv^{-5/2} _i \sqrt {a\over c},\ x={u\over v_i} ,\ \theta =
{v_0^2\over v_i^2},\ \mu = {c\over b},\ 
\epsilon^2={m\over M} \eqno(21)$$
(19) can be rewritten in the dimensionless form $$
\phi = {\omega +x^{5/2}\over x^3}\left [ \sqrt{6\over \pi}{
\epsilon ^2\over \omega \mu }(x^2 -1)(\omega +2x^{5/2})+
(x^2 -\Theta)/2 \right ].\eqno(22)$$    
\medskip
It is obvious from (22) that $\phi $ is not monotone in $x$ and
therefore $T(E)$ is not single-valued if the parameter $\omega$ is big
enough. In a Helium plasma where $M/m \approx 7000,$ we take $\
c=4\cdot 10^{13}cm^6 sec^{-4},\ \theta =10$ and find that $\omega $
should be larger than about 113 in order to have a transition
corresponding to an S-shaped $T(E)$ curve; see Fig.1, where $\omega
=250$. For smaller $c$ the critical value of $\omega$ is lower.

In terms of the ion temperature and the degree of ionization
$\omega$ can be represented as$$
\omega =\pi L{e^4\over \sigma (kT_i)^4}{n\over N}.$$
This is equal approximately $2\cdot 10^6 n/N$ for Argon with the
ambient temperature of the background gas. We see that the transition
requires $n/N \sim 10^{-4}$ or higher.  Real values for the electric
field and current density $j$ can be found with the help of following
relations:$$ El=1.28\cdot 10^{-4}\left (T_i\over 300\right
)\sqrt{\omega \phi},
\ \ jl=2.6\cdot 10^{-3}\left (T_i\over 300\right )^{5/2}{x^3\sqrt
{\omega^3\phi}\over \omega +x^{5/2}}.$$ Here $l$ is in $cm$, $E$ and
$j$ are in $volts/cm$ and $mA/cm^2$ respectively.
\bigskip\noindent
\centerline{\bf The T = const Approximation}
\medskip
As a first step in solving the nonuniform case (12-14) let us assume
that the electron temperature is constant, $u(r)=u(0)$. It seems
reasonable to study such a simplified problem both as a guide for the
more general case which we shall consider later and as an
approximation which often yields reliable results for the positive
plasma column$^6$.
\medskip
The energy balance equation (14) cannot hold now for all $r\leq R,$
but at the point $r=0$ it reduces to (19), and our task is to solve
n(12,13) with the conditions (16--18). {}For fixed E and $n=n(0)$
Eq.(19) determines $u$ and therefore the electron temperature in the
tube. We then integrate (12,13), find the profiles $n(r),\ w_r(r)$,
compute $w_z(r)$ with the help of (11), and using (20) find the total
current $I(E)$, which gives the current-voltage characteristic of the
plasma. We can neglect the difference between $b$ and $g$, because in
noble gases $g/b-1\sim \mu \sim 10^{-4}$.
\medskip
To solve Eq.(12,13) numerically we pick a value of $u$ and find a
suitable $n(0)$ which allows to satisfy (16).  Eq. (19) is used for
the calculation of $E$ for each choice of $u$ and $n(0)$. In this way
we obtain the functions $n(r),\ w_z(r)$ and substitute them into
(20). {}Fig.2 shows the electron temperature and the electric current
thus computed versus $E^2$ in relative units when $l/R =0.2.$ Curves
$T(E)$ are single-valued for $\omega\leq 66$ with the same $\Theta ,c$
as before in a Helium plasma.
\medskip
\bigskip\noindent
\centerline {\bf T$\not \equiv $ const}
\medskip
We solved numerically Eqs. (12)--(14) with boundary conditions
(16--19). Technically we choose for each $u(0)$ a trial value of
$n(0)$, compute E with the help of Eq.(19), and solve the differential
equations (12--14) for this triple. The resulting $w_r(r)$ will
generally not satisfy the boundary condition (16) at $r=R$ and we then
iterate with a different $n(0)$. This search can be easily optimized,
the procedure usually converges very fast and yields the profiles of
$n(r)$ and $u(r)$ for $r\leq R$.  {}From these we can obtain $F(r),\
w_z(r),\ w_r(r)$ for a given external field E. In addition we also
find the total current and the mean electron temperature over the tube
cross section (20). The parts of the $T(E)$ and $I(E)$ curves with
negative derivatives are unstable and physically
non-accessible$^{4,5}$, which also shows up in the computation being
extremely unstable there when we try to find the solutions
numerically. Thus for $E=E'$ in {}Fig.3 and any trial temperature on the
tube axis inside the gap region (0.046-0.065 V), $T(r)$ immediately
jumps to a point corresponding to $T'$ after a few steps of
integration. This difficulty does not occur when we keep $T=const$.
\medskip
A qualitative description of the results is as follows:
\medskip\noindent
a) $n(r)$ always decreases monotonically. {}For some range of parameters
a kinetic transition takes place, $n(0)$ is then not a single-valued
function of the external field or the mean electron temperature. The
decrease of the volume recombination rate $c$ enhances the transition.

\medskip\noindent
b) The I-V characteristic for the total current I(E) and the mean
electron temperature show transition like behavior for $\omega >
60$. This is close to that obtained for the model $T=const$.
\medskip\noindent
c) $T(r)$ is a smooth monotonically decreasing curve when $E$ is
large; for very small $E$ the electron temperature passes a maximum at
some $r<R$. This maximum is caused by the electron heating in the
ambipolar radial field $F$ which is equal to zero on the tube axis.
At stronger fields $E$ this effect is disguised by the external field
producing more heat near the axis where the electron density is
higher.
\medskip\noindent
d) The radial drift of electrons and ions as well as the ambipolar
electric field $F(r)$ rise strongly near the tube walls, the latter
even goes to infinity in this model but the wall potential with
respect to the tube axis is finite as one can see from Eq.(8).
\medskip
We present a few curves for $\omega =90$ in Figs.3,4 which illustrate
the transition, the temperature profiles across the tube cross section
and the radial dependence of the transverse electric field $F(r)$. As
$E$ is varied we expect jumps in $T(E)$ and $I(E)$ at the arrows in
{}Fig.3.
\bigskip\noindent
{\bf IV. Discussion}
\medskip
In this work we investigated properties of the stationary state of a
partially ionized plasma in a cylindrical tube maintained by a source
of ionization and subjected to an external axial electric field
$E$. Our calculations were done in a collisional two-fluid scheme. The
neutrals are treated as a given background with fixed density $N$ and
temperature $T_i$ while the ions were assumed to have the same uniform
temperature $T_i$ and a density $n(r)$ equal to that of the electrons,
i.e. charge neutrality in the interior of the tube.

The assumption that the heating of the electrons by the external and
ambipolar fields together with the injection of relatively warm
electrons in the process of ionization does not heat the neutrals and
ions very much seems reasonable when one takes into account the
density and heat conductivity of the neutrals and the strong coupling
between them and the ions. To see this in a semiquantitative way we
note that the temperature distribution for the neutrals is governed by
the equation$$ {1\over r}{d\over dr}\left (r{dT\over dr}\right
)=-{P(r)\over \lambda}
\eqno(23)$$   
where $P(r)$ describes the heating sources and $\lambda$ is the
thermal conductivity which can be taken to be constant. By keeping the
walls of tube at a given temperature $T_0$ we impose the boundary
condition $T(R)=T_0$ which gives$$ T(r)=T_0 +{1\over
\lambda}\int_r^R{dx\over x}\int_0^x yP(y)dy.\eqno(24)$$ Let us
consider three cases: 1) $P(r)$ is a constant, 2) linearly falls to
zero, and 3) approaches zero parabolically. Then letting$$ {\bar P}=
{2\over R^2}\int_0^R P(r)rdr$$ be the average intensity of the heat
source, Eq.(24) yields, on the tube axis$$ T_1(0)=T_0+{1\over 4}{{\bar
P}R^2\over \lambda},\ T_2(0)=T_0+{5\over 12}{{\bar P}R^2\over
\lambda},\ T_3(0)=T_0+{3\over 8}{{\bar P}R^2\over
\lambda}$$
respectively. The condition$$ {\bar P}<<2\lambda T_0/R^2\eqno(25)$$
then guarantees an approximately uniform temperature of the neutrals.
In fact we take $T_0 =T_i$ and Eq.(25) can be rewritten in the form$$
E{\bar j}<<{1\over \sigma R^2}\sqrt{3\over M}(kT_i)^{3/2}$$ if, for
simplicity, we neglect the effects of the neutral gas non-ideality and
only take into account the heating by the external field, $P=Ej$. {}For
a typical example of the situation considered in this paper, which
will be described later in this section, $E{\bar j}$ is four orders of
magnitude smaller than the right hand side of (25).

The assumption of a constant ion temperature will hold only
approximately. We expect that away from the tube axis the ion
temperature will become somewhat larger than the temperature of
neutrals but not significantly so, except maybe in close vicinity of
the wall$^{6,7,8}$. Since the transition we are studying here takes
place when $T$ is only a few times higher than that of the neutrals,
we may assume a constant ion temperature $T_i$, at least for the
majority of ions.

The kinetic model on which the hydrodynamic equations (6--10) are
based is the nonlinear Boltzmann equation (2) for the electron
distribution function $f(r,{\bf v},t)$. To obtain the fluid equations
we have used the approximation of replacing $f$ by $M_f$ in evaluating
the integrals for the time derivatives of the hydrodynamic variables.
The justification given in section 3 is the dominant effect of e-e
collisions.  But, as is clear from the behavior of the e-e collision
cross section, which decreases for large speeds as $v^{-4}$, $f(v)$
will not be uniformly close to $M_f(v)$ for all $v$.  {}For large
speeds, the behavior of $f$ in the stationary state will be dominated,
for $|E|$ not very small, by collisions with neutrals which lead to a
Dryuvesteyn type tail of the form$^{14}$ $f \sim e^{-Av^4 /E^2}.$ Very
slow electrons, on the other hand, are strongly influenced by the e-i
interactions and their distribution is distorted near $v=0$
also. However, since we are only interested in the evaluation of
averages of quantities which are not too peaked at very slow or very
high speeds, the main contribution will come from typical values of
${\bf v}$, $v \sim u =
\sqrt {3kT/m}$, where
the bulk of electrons are. In this region of velocity space $f$ will
be close to $M_f$ whenever the ``relaxation time'' $\tau _{ee}$
associated with e-e collisions is small compared with other relaxation
times in the problem which would try to drive the electron
distribution away from a local Maxwellian. We now evaluate the
different relaxation times for typical electrons in the spirit of
plasma kinetic theory$^{11,12}$.
\medskip
The frequency of e-e collisions$^1$ is $bn/u^3$ so $\tau_{ee}$, the
time of energy and momentum relaxation in such collisions, can be
approximated as $$\tau _{ee} \approx {u^3 \over bn}.\eqno(26)$$ The
relaxation time of the newly created electrons, whose mean speed $v_0$
is higher than $u$, can be obtained from (26) by replacing $u$ with
$v_0 = \Theta ^{1/2} u$.  On the other hand an electron with speed $v$
loses, in a collision with a slow ion having velocity $u_i$, a
fraction $2m/M$ of its energy.  Such collisions take place with
frequency, $\nu_{ei} = bn/u^3 = {\tau_{ee}}^{-1}$, so the e-i energy
relaxation time is $$
\tau _{ei} \sim {M\over m}\tau _{ee} .\eqno(27)$$
By the same reasoning, the energy relaxation time with neutrals, where
the collision frequency $\nu_{en}$ is $v/l$, is
$$
\tau _{en}\approx M/m \nu_{en} = {Ml/mu}.\eqno(28)$$
\medskip 
 We thus always have $\tau_{ee}/\tau_{ei} = m/M <<1$ and the
requirement that $\tau_{ee}/\tau_{en} << 1$ is just the left side
inequality in (1a).  This assures that e-e collisions will play a
dominant role in determining the energy distribution of the electrons,
at least for the case of a uniform plasma with fixed electron density
$n$, considered in Refs.3-5. In the present case we also have to
consider the electron lifetimes relative to recombination in the bulk,
$\tau_B$, and on the wall $\tau_W$.  A simple calculation yields
$$
\tau _B = {u^3\over cn}={b\over c}\tau _{ee} \approx 10^4 \tau_{ee}.
\eqno(29)$$
To estimate $\tau _W$ we compare the flux of neutralizing ions to the
walls, $u_i n(R)\cdot 2\pi R$, and the total number of electrons in
the tube cross section $\bar n\pi R^2$: $\tau _W \approx R\bar n/2u_i
n(R).$ Taking roughly $n(R)/\bar n \sim 1/20$ we obtain, after some
substitutions, $$
\tau _W \approx 10 R/u_i \approx 10(R/l)(MT/mT_i)^{1 \over 2}
\tau_{en}.\eqno(30)$$
Thus for $R/l \geq 100$, $\tau_W \geq \tau_{en}$ and the arguments
used for fixed $n$ should apply.

We note here however that in order to get $M \sim M_f$ with {\it both}
$T$ and ${\bf w}$ determined by the hydrodynamic equations, one
actually needs to assume also that $\tau_{ee}$ is small compared to
the momentum relaxation times, i.e.\ that $(\tau_{ee} \nu_{ei})$ and
$(\tau_{ee} \nu_{en})$ are small.  This was assumed for the simplified
model equations$^5$ but clearly does not hold in the plasma.  A better
approximation scheme to solving (2) would therefore be to take $f \sim
M^0_f$, a local Maxwellian with mean drift ${\bf w} = 0$. The leading
corrections to this Maxwellian would be of order $\sqrt {m/M}$, and
would have to be computed from a linearization of (2) about
$M^0_f$. The result would then be the input for the self-consistent
evaluation of $T$.  This is what was essentially done heuristically
and approximately in Ref.4. To make the scheme work rigorously the
effective strength of $E$ and hence $w$ must also be of this order.
This is indeed what we find here, for the range of $E$ in {}Figs. 1--3,
with $w/u \sim \sqrt {m/M}$. In fact the results obtained in Ref.4 are
qualitatively similar to those obtained with the present approximation
scheme for uniform systems, which facilitates computation; see also
Appendix.
\bigskip 
\centerline {\bf Realizability of Transition}
\medskip

We shall consider now, within the scheme of section 3 and the even
simpler relaxation time estimates discussed above, the physical
conditions necessary for observing the kinetic transition manifested
in the $S$-shaped curves in {}Figs. 1--3.  Since the origin of this
behavior is a changeover from e-i to e-n collisions as the dominant
mechanism for the dissipation of the energy which electrons gain from
the external field, it is necessary that at the initiation of the
transition, i.e. at the bottom part of the $S$-curve, the electron
temperature, $T_b$, should be low enough for $\nu_{ei}/\nu_{en}$ to be
large.  The requirement that this ratio, which can be written as $(4
\pi n L/9N) (e^4/\sigma)/(kT_b)^2$, be large compared to one and that
we still have $n/N << 1$ imposes a strong constraint on the
temperature of the electrons $T_b$.  Assuming a Helium plasma with the
neutrals and ions at room temperature, $T_i = 300^0 K$, $R = 1 cm$,
$\theta = (v^2_0/v^2_i) = 10$, corresponding to an energy of 1/4 eV
for newly created electrons, $\sigma = 5\times 10^{-16}\ cm^2$ (see
Ref.17) the ionization fraction at the center of the tube,
corresponding to the bottom in {}Fig. 3 is, $n/N \sim 4\cdot 10^{-5}$.
We took $l/R=0.2$ which implies $N\approx 10^{16}cm^{-3}$
corresponding to $NkT_i = 10^{-1}\ Torr$ and $n(0)\sim 4\cdot 10^{11}
cm^{-3}.$ The electric field in {}Fig.3 goes from zero to about $0.2\
V/cm$ and the total current does not exceed $4 mA$.  {}For a Neon plasma
$\sigma$ is smaller by a factor of 9 and so is the requirement on the
ionization fraction.

Whether conditions like these can be achieved experimentally is at the
moment an open question.  What is clear however is that if we have to
rely on ionization caused by the external field $E$ then $T_b$ will be
about 1{}~eV and the electrons will never be coupled strongly to the
ions for $n/N << 1$.  To have a chance of seeing a transition
corresponding to a crossover from e-i to e-n coupling we would have to
consider an almost fully ionized plasma in which the remaining
neutrals would prevent the runaway effect caused by the external field
$E$ when only e-e and e-i collisions are considered, (see Ref.11).

To gain a better understanding of the dynamics involved in this
non-equilibrium kinetic transition, let us analyze, in simple physical
terms, the stationary state of our system in the presence of the
electric field ${\bf E}$. We assume as before that in each e-i or e-n
collision the fraction of energy lost is $2m/M$ and the direction of
motion of the colliding electron is randomized. Under the action of
the force $-e{\bf E}$ an electron between two successive collisions
changes its velocity from ${\bf v}$ to ${\bf v}-e{\bf E}\tau /m$,
where the mean time of the free flight, is$$
\tau = \left ({bn\over v^3} + {v\over l}\right )^{-1}.\eqno(31)$$
When $v$ corresponds to the thermal velocity $u$ the drift velocity 
of electrons obtained from (31) is$$
{\bf w}\approx -{e{\bf E}\over 2m}\left ({bn\over u^3} + 
{u\over l}\right )^{-1} .\eqno(32)$$
(This almost coincides with Eq.(13) for $w_z$). In the stationary 
state the average energy gain between two collisions$$
\Delta W = {(eE\tau )^2\over 2m}$$
should be equal to the average energy loss $(m/M)m(u^2 -v^2_i)$,$$
\left ( {eE\over m}\right )^2 ={2m\over M}(u^2-v_i^2)\left ({bn\over 
u^3} +{u \over l}\right )^2 .\eqno(33)$$ We see from (33) that $u(E)$
is not a monotone function when $q=bnl/v_i^4$ is larger than about 12.
This gives in a simplified form the origin of the kinetic transition.
{}For this analysis to be reasonable the energies of the electrons must
not be spread out too much, i.e.\ the bulk of electrons must have
their speeds close to $u$. The e-e collisions provide the nonlinear
cooperative coupling which brings about this condition. In the
cylindrical plasma confined in a tube which was studied here the
critical value of $q$ is found very close to 10 for plasma parameters
used in {}Figs.1--3.

An S-shaped $T(E)$ or in other words a non-monotone behavior of the
function $E(u)$ is caused by the decreasing term $bn/u^3$ in (33). The
recombination of electrons near the tube walls makes the density $n$ a
decreasing function of $u$, due to the growth of the charged particle
mobility with temperature and therefore enhances the observation of
our transition. The opposite role is played by the volume
recombination whose rate decreases with the electron temperature ($u$
here) and leads to a rise of density with $u$. We thus have a
competition between these processes in the cylindrical geometry and
may anticipate that the smaller is the bulk recombination the easier
would be an experiment.
\bigskip
{\bf Acknowledgments}

We thank Dr.\ Robert Barker both for useful specific comments and for
general encouragement during the course of this work.  We also thank
Dr. Spencer Kuo for looking at possible experimental realizations of
our kinetic transition and the referee for many helpful comments.
Work supported by Air {}Force Office of Scientific Research Grant 0159
4--26435.
\vfill
\eject
\centerline {\bf Appendix: Nature of Approximations}
\medskip
We rewrite now Eq.(2) in a dimensionless form in order to better see
the mathematical nature of the approximations made in our work.
Setting $\epsilon =
\sqrt{m/M},{}~t' =\epsilon tv_i /l,\ {\bf v' }= {\bf v}/v_i,\ {\bf r'}=
{\bf r}/R,\ \epsilon {\bf w' }={\bf w}/v_i,\ \delta = l/R,  {\bf E'}=
(el/\epsilon mv_i^2){\bf E},\ {\bf F' }=(el/\epsilon mv_i^2){\bf F }, 
\ \lambda = l\sqrt{ac/v_i^5},\ n(r')=n(r)\sqrt{c/av_i^3}$, 
and dropping primes we get$$
\epsilon {\partial f({\bf r},{\bf v},t)\over \partial t}-\epsilon 
({\bf E +F})\cdot {\bf \nabla _v}f + \delta ({\bf v\cdot \nabla _r}f)=$$
$$\lambda (\psi -{n\over x^3}f) +\epsilon ^2 {1\over v^2}{\partial \over
\partial v}\left [ (4\omega n/9+ v^4)\left ({\bar f} + 
{1\over 3v}{\partial{\bar f}\over \partial v}\right ) \right ]+ 
{4\omega n\over 9 v^3}{\hat L}f +v({\bar f}-f) +Q[f],\eqno(A1)$$
where $\omega$ is defined in (21) and we have$$
\int \psi d^3 v=1,\quad \int {\bf v}\psi d^3 v=0,\quad 
\int v^2\psi d^3 v=\Theta ,$$
$$\int f d^3 v=n(r),\quad \int {\bf v}f d^3 v=\epsilon
n(r){\bf w}(r),\quad \int v^2f d^3 v=n(r)
[x^2 (r)+{\epsilon}^2 w^2 (r)].$$

We restrict ourselves now to the homogeneous case ($R=\infty $) with a
fixed electron density, $n,$ without ionization - recombination.
Eq.(A1) then takes the form$$ {\partial f\over \partial \tau}-{\bf
E}\cdot {\bf
\nabla _v}f ={\epsilon \over v^2}{\partial \over
\partial v}\left [ (q+ v^4)\left ( {\bar f} + 
{1\over 3v}{\partial {\bar f}\over \partial v}\right ) \right ]+ 
{1\over \epsilon}\left \{ {q\over v^3}{\hat L}f +v({\bar f}-f) +
Q[f]\right \} .\eqno(A2)$$

In the stationary state the equations for the moments have the form
$${\bf E}+{1\over \epsilon}\int {\bf v} 
\left ({2q\over v^3}+v \right )\Phi ({\bf v} )d^3 v =0,\eqno(A3)$$
$$ Ew +{4\pi q\over 3}\Phi (0)-\int \left ({q\over v} +v^3 -{4v\over
3}\right )\Phi ({\bf v} )d^3 v =0,\eqno(A4)$$ where $q=bnl/v^4_i$ as
before and we have set $\Phi =f/n$ so that$$
\int {\bf v}\Phi d^3 v =\epsilon {\bf w},\quad
\int v^2 \Phi d^3 v = x^2 + {\epsilon}^2 w^2.$$
If one uses the substitution (7), which now has the form,
$$
\Phi ({\bf v})\to (3/2\pi )^{3/2}x^{-3}\exp\left [-{3\over 2x^2} 
({\bf v} -\epsilon {\bf w})^2 \right ] \eqno(A5)$$ and neglects
${\epsilon}^2(w/x)^2$ compared to unity Eqs.(A3,A4) transform into the
set$$ E= {2w\over x^3}\sqrt{6\over \pi}(q+4x^4/9 )=0,\eqno(A6)$$
$$Ew=\sqrt{6\over \pi}{x^2 -1\over x^3}(q+8x^4/9 )=0.\eqno(A7)$$
In the stationary case (A6,A7) can be solved for $w(E)$
and $x(E)$ yielding non-unique solutions, when $q>7.6$ (see Ref.4).
The solution of (A6,A7) for $x$ can be found from the relation$$
E^2 ={12\over \pi}{x^2-1\over x^6}(q+4x^4/9 )
(q+8x^4/9 ).\eqno(A8)$$
Eq.(A8) is close to Eq.(33), which was obtained in a rough
approximation without referring to the kinetic equation.
\medskip
It is natural to expand $f({\bf v},t)$ for the spatially homogeneous
case in a series in $\epsilon$
$$f({\bf v},t)=\sum _{j=0} f_j ({\bf v},t)\epsilon ^j .\eqno(A9)$$
Substituting (A9) into (A2) gives a 
set of coupled equations for $f_j$ where higher
components can be expressed through lower ones.
We do not use this method in the present work, instead we have solved
Eq.(A1) using the method of moments with $\epsilon =1.4\times 10^{-4},
\ \delta =0.2, \ \lambda \approx 0.1,$ and $\Theta =10.$ In order to
have a closed set of differential equations we have taken$$
f_1=\left ({3{\bf v\cdot w}\over x^2}
\right ){\bar f}$$
which comes from the shifted Maxwellian. We expect the effect of our
additional approximation to be small.
\vfill
\eject
\centerline {\bf REFERENCES}
\bigskip
\item {[1]}\ F.F.Chen, {\it Introduction to Plasma Physics} (Premium
Press, New York, 1974), pp. 139-140, 158-160.
\bigskip
\item {[2]}\ V.A.Godyak, R.B.Piejak, and B.M.Alexandrovich,
Plasma Sources Sci. Technol. {\bf 3}, 169 (1994);

\item {[3]}\ IEEE Conference Record -- Abstracts, 1995 IEEE International
Conference on Plasma 

Science.
\bigskip
\item {[4]}\ A.V.Rokhlenko, Phys.Rev. A {\bf 43}, 4438 (1991).
\bigskip

A.V.Rokhlenko and J.L.Lebowitz, Phys.Fluids B {\bf 5},
1766 (1993).
\bigskip
\item {[5]}\ E.Carlen, R.Esposito, J.L.Lebowitz, R.Marra, and A.Rokhlenko,
Phys.Rev. E, {\bf 52}, R40 (1995).
\bigskip
\item{[6]}\ R.N.Franklin,\ {\it Plasma Phenomena in Gas Discharges}
(Clarendon Press, Oxford, 1976), pp.20-37;
\bigskip
\item {[7]}\ K.-B.Persson, Phys.Fluids {\bf 5}, 1625 (1962).

D.B.Ilic, J.Appl.Phys. {\bf 44}, 3993 (1973).
\bigskip
\item {[8]}\ H-B Valentini, J.Phys.D {\bf 21}, 311 (1988).
\bigskip
\item {[9]}\ U.Kortshagen, Phys.Rev. E {\bf 49}, 4369 (1994);

V.I.Kolobov and W.N.G.Hitchon,\ Phys.Rev. E {\bf 52}, 972 (1995).
\bigskip
\item {[10]}\ V.E.Golant, A.P.Zhylinsky, and I.E.Sakharov, {\it Fundamentals
of Plasma Physics} (Wiley, New York, 1980).
\bigskip
\item {[11]}\ R.Balescu, {\it Transport Processes in Plasmas} (North-Holland,
Amsterdam-Oxford-New York-Tokyo, 1988), pp. 113, 138, 775-786.
\bigskip
\item {[12]}\ E.M.Lifshitz and L.P.Pitaevsky, {\it Physical Kinetics}
(Pergamon, New York, 1981), pp. 93, 172, 181-184;

A.von Engel,\ {\it Ionized Gases} (AIP Press, New York, 1993),
pp. 29-30, 243, 292;

C. Brown,\ {\it Basic Data of Plasma Physics} (AIP Press, New York, 1993).
\bigskip
\item {[13]}\ N.J.Carron, Phys.Rev. A {\bf 45}, 2499 (1992).
\bigskip
\item {[14]}\ M.J.Druyvesteyn, Physica {\bf 10}, 61 (1930);

M.J.Druyvesteyn and E.M.Penning, Rev.Mod.Phys. {\bf 12}, 87 (1940).
\bigskip
\item {[15]}\ L.G.Christophorou, {\it Electron--Molecule Interactions
and Their Applications, vol.2} (Academic Press, Orlando, 1984),
pp. 65-88.

A.J.Cunningham and R.M.Hobson, Phys.Rev. {\bf 185}, 98 (1969).
\bigskip
\item {[16]}\ R.Nagpal and A.Garscadden, Phys.Rev.Lett. {\bf 73},
1598 (1994).
\bigskip
\item {[17]}\ L.G.H. Huxley and R.W. Crompton, {\it The Diffusion and 
Drift of Electrons in Gases} (Wiley, New York, 1974), Chapter 14.

\vfill
\eject

\centerline {\bf FIGURE CAPTIONS.}
\bigskip
\centerline {Fig.1}
\medskip
The dependence of $x$ (representing the electrons thermal speed) and
the current density on $\phi$ (the electric field squared) for the 
spatially homogeneous case $(R=\infty )$ when $\omega =250.$ The 
dimensionless units are defined in Eq.(21) in the text.
\bigskip
\centerline {Fig.2}
\medskip
The electron temperature $T$ and total current $I\cdot R$ (in
$mA\cdot cm$) vs the electric field $E\cdot R$ (in volts) when 
$T$ is assumed constant in the tube cross section,
$T_i=300^0$ and $\omega =90.$
\bigskip
\centerline {Fig.3}
\medskip
Plots of the mean electron temperature and total current when 
$T\not \equiv const$ and $\omega =90.$ (The same units as in Fig.2).
The hysteresis loop is indicated by arrows. At values of the field
between the end points of the loop, like $E'$, the computation 
leads to values on the lower or upper branches of the loop, 
determined by how close the starting point is to one of them.
\bigskip
\centerline {Fig.4}
\medskip
The radial profiles of the mean electron speed and ambipolar
electric field $F(r)\cdot R$ (in volts) when $\omega =90$ for two
regimes: the solid lines correspond to a small field 
$(E\cdot R\approx 0.04 V,\ {\bar T}/T_i \approx 2.5),$ the dotted 
lines to a larger field $(E\cdot R\approx 0.07 V,\ {\bar T}/T_i
\approx 8).$

\end